\documentclass[10pt, conference, compsocconf]{IEEEtran}
\ifCLASSINFOpdf
  \usepackage[pdftex]{graphicx}
  \DeclareGraphicsExtensions{.pdf,.jpeg,.png}
\else
\fi
\usepackage{color}
\usepackage{xspace}
\usepackage[normalem]{ulem}
\usepackage[font=footnotesize]{subfig}
\usepackage[hidelinks]{hyperref}
\usepackage{cleveref}
\Crefname{figure}{Fig.}{Figs.}
\Crefname{table}{Tab.}{Tabs.}
\Crefname{equation}{Eqn.}{Eqns.}
\Crefname{algorithm}{Algo.}{Algos.}
\Crefname{section}{Sec.}{Secs.}
\Crefname{lstlisting}{List.}{Lists.}
\usepackage{booktabs}
\usepackage{csquotes}
\usepackage[super]{nth}
\usepackage{csquotes}
\usepackage{xspace}
\usepackage[binary-units=true]{siunitx}
\usepackage{wrapfig}
\usepackage[nolist]{acronym}

\makeatletter
\DeclareRobustCommand\onedot{\futurelet\@let@token\@onedot}
\def\@onedot{\ifx\@let@token.\else.\null\fi\xspace}
\def\etal{\textit{et~al}\onedot~}
\def\eg{e.g\onedot, } 
\def\ie{i.e\onedot, } 
 
\def\etc{etc\onedot} 

\makeatother

\def\clap#1{\hbox to 0pt{\hss #1\hss}}%
\def\initials#1{\protect\clap{\smash{\raisebox{1.4ex}{\tiny{\textsf{\textit{#1}}}}}}}%
\makeatletter
\newcommand{\EDIT}[4][]{\protect\@ifundefined{hidecomments}{%
  \strut{\color{#3}{\hspace{0pt}\initials{#2}\protect\sout{#1}{~#4}}}%
  }{}}
\newcommand{\NOTEboxed}[3]{\protect\@ifundefined{hidecomments}{%
  {\begin{center}\fbox{\parbox{0.97\linewidth}{\protect\EDIT{#1}{#2}{#3}}}\end{center}}
  }{}}
\makeatletter
\newcommand{\DefAuthor}[2] 
{%
  \expandafter\newcommand\csname #1edit\endcsname[2][]{\protect\EDIT[##1]{#1}{#2}{##2}}
  \expandafter\newcommand\csname #1\endcsname[1]{\protect\csname #1edit\endcsname{[##1]}}
  \expandafter\newcommand\csname #1boxed\endcsname[1]{\NOTEboxed{#1}{#2}{##1}}
}
\makeatother

\newcounter{AspectCNT}%
\renewcommand{\theAspectCNT}{Aspect-\arabic{AspectCNT}}%
\newcommand{\Aspect}[1]{%
  \refstepcounter{AspectCNT}
  \theAspectCNT
  \label{#1}}

\newcounter{TaskCNT}%
\newcommand{\inlineimage}[1]{\begingroup
\setbox0=\hbox{\includegraphics[height=0.75\baselineskip]{#1}}%
\parbox{\wd0}{\box0}\endgroup}

\newcommand{\inlineimagetitle}[2]{\begingroup
\setbox0=\hbox{\includegraphics[height=#1]{#2}}%
\parbox{\wd0}{\box0}\endgroup}

\usepackage{listings}
\usepackage{xcolor}
\colorlet{punct}{red!60!black}
\definecolor{background}{HTML}{EEEEEE}
\definecolor{delim}{RGB}{20,105,176}
\colorlet{numb}{magenta!60!black}

\lstdefinelanguage{json}{
    basicstyle=\tiny\ttfamily,
    numbers=none,
    linewidth=\linewidth,
    rulecolor=\color{background},
    fillcolor=\color{background},
    numberstyle=\scriptsize,
    stepnumber=1,
    numbersep=8pt,
    showstringspaces=false,
    breaklines=true,
    frame=tbrl,
    backgroundcolor=\color{background},
    literate=
     *{0}{{{\color{numb}0}}}{1}
      {1}{{{\color{numb}1}}}{1}
      {2}{{{\color{numb}2}}}{1}
      {3}{{{\color{numb}3}}}{1}
      {4}{{{\color{numb}4}}}{1}
      {5}{{{\color{numb}5}}}{1}
      {6}{{{\color{numb}6}}}{1}
      {7}{{{\color{numb}7}}}{1}
      {8}{{{\color{numb}8}}}{1}
      {9}{{{\color{numb}9}}}{1}
      {:}{{{\color{punct}{:}}}}{1}
      {,}{{{\color{punct}{,}}}}{1}
      {\{}{{{\color{delim}{\{}}}}{1}
      {\}}{{{\color{delim}{\}}}}}{1}
      {[}{{{\color{delim}{[}}}}{1}
      {]}{{{\color{delim}{]}}}}{1},
}

\newcommand{\belowfig}{\vspace{-0.25cm}}

\begin{document}
%
\title{\inlineimagetitle{1cm}{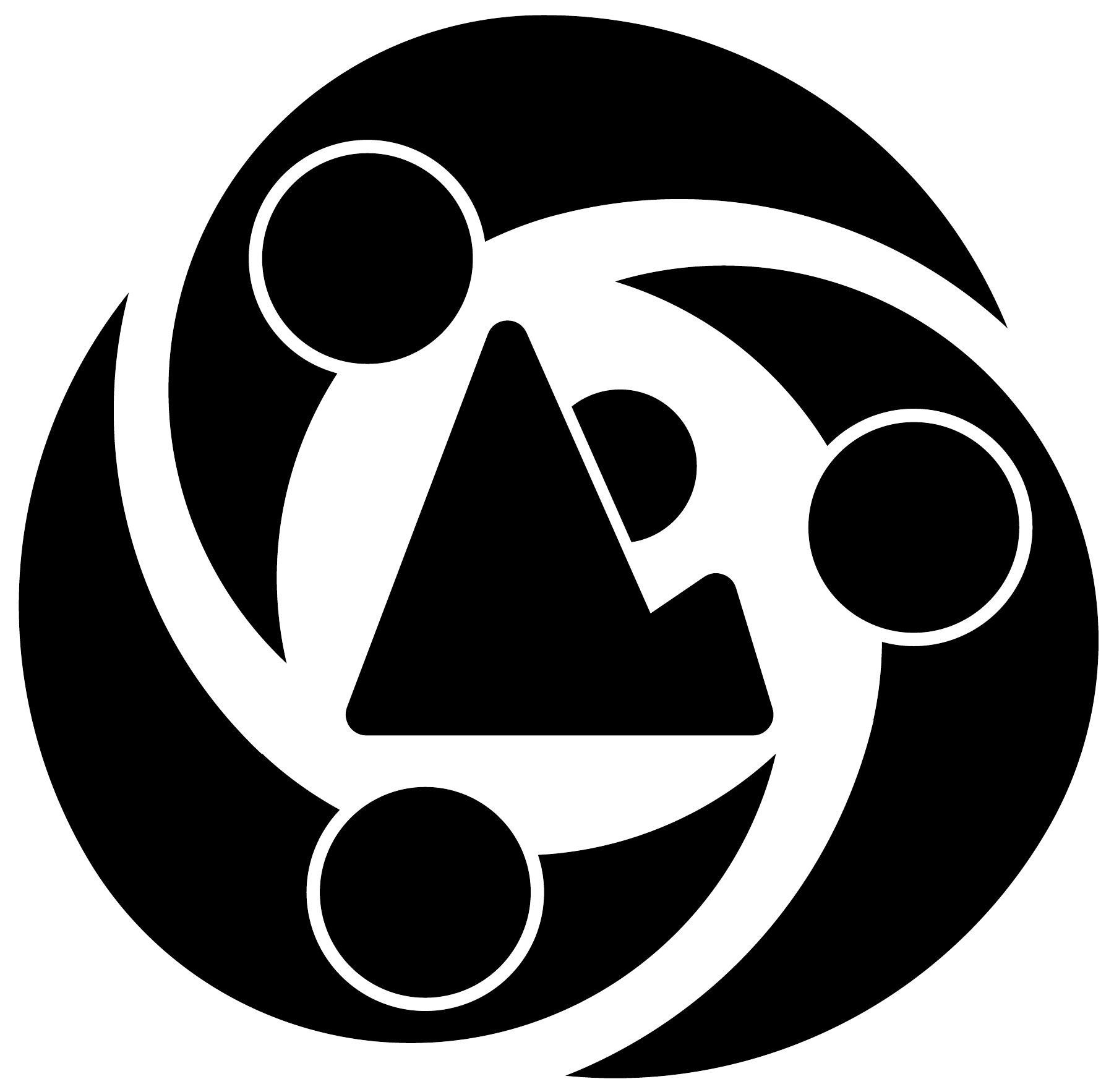} COLiER: Collaborative Editing of Raster Images}

\author{
\IEEEauthorblockN{Ulrike Bath, Sumit Shekhar, J\"urgen D\"ollner, Matthias Trapp}
\IEEEauthorblockA{Hasso Plattner Institute, Faculty of Digital Engineering, University of Potsdam, Germany\\
ulrike.bath@student.hpi.de, sumit.shekhar@hpi.de, juergen.doellner@hpi.de, matthias.trapp@hpi.de}
}


%


\maketitle

\begin{abstract}
Various web-based image-editing tools and web-based collaborative tools exist in isolation. 
Research focusing to bridge the gap between these two domains is sparse. 
We respond to the above and develop prototype groupware for real-time collaborative editing of raster images in a web browser. 
To better understand the requirements, we conduct a preliminary user study and establish communication and synchronization as key elements.  
The existing groupware for text documents, presentations, and vector graphics handles the above through well-established techniques. 
However, those cannot be extended as it is for raster graphics manipulation. 
To this end, we develop a document model that is maintained by a server and is delivered and synchronized to multiple clients.
Our prototypical implementation is based on a scalable client-server architecture: using WebGL for interactive browser-based rendering and WebSocket connections to maintain synchronization. 
We evaluate our work qualitatively through a post-deployment user study for three different scenarios.
\end{abstract}


\definecolor{dkgreen}       {rgb}{0.0,0.5,0.0}
\DefAuthor{MT}{dkgreen}

\definecolor{dkred}       {rgb}{0.5,0.0,0.0}
\DefAuthor{SSh}{dkred}

\definecolor{dkgelb}       {rgb}{0.3,0.01,0.06}
\DefAuthor{UB}{dkgelb}

%
\IEEEpeerreviewmaketitle

\section{Introduction}

\begin{figure}[tb]
\centering
\subfloat[Madonna in Floral Wreath]{%
 \includegraphics[height=3.55cm]{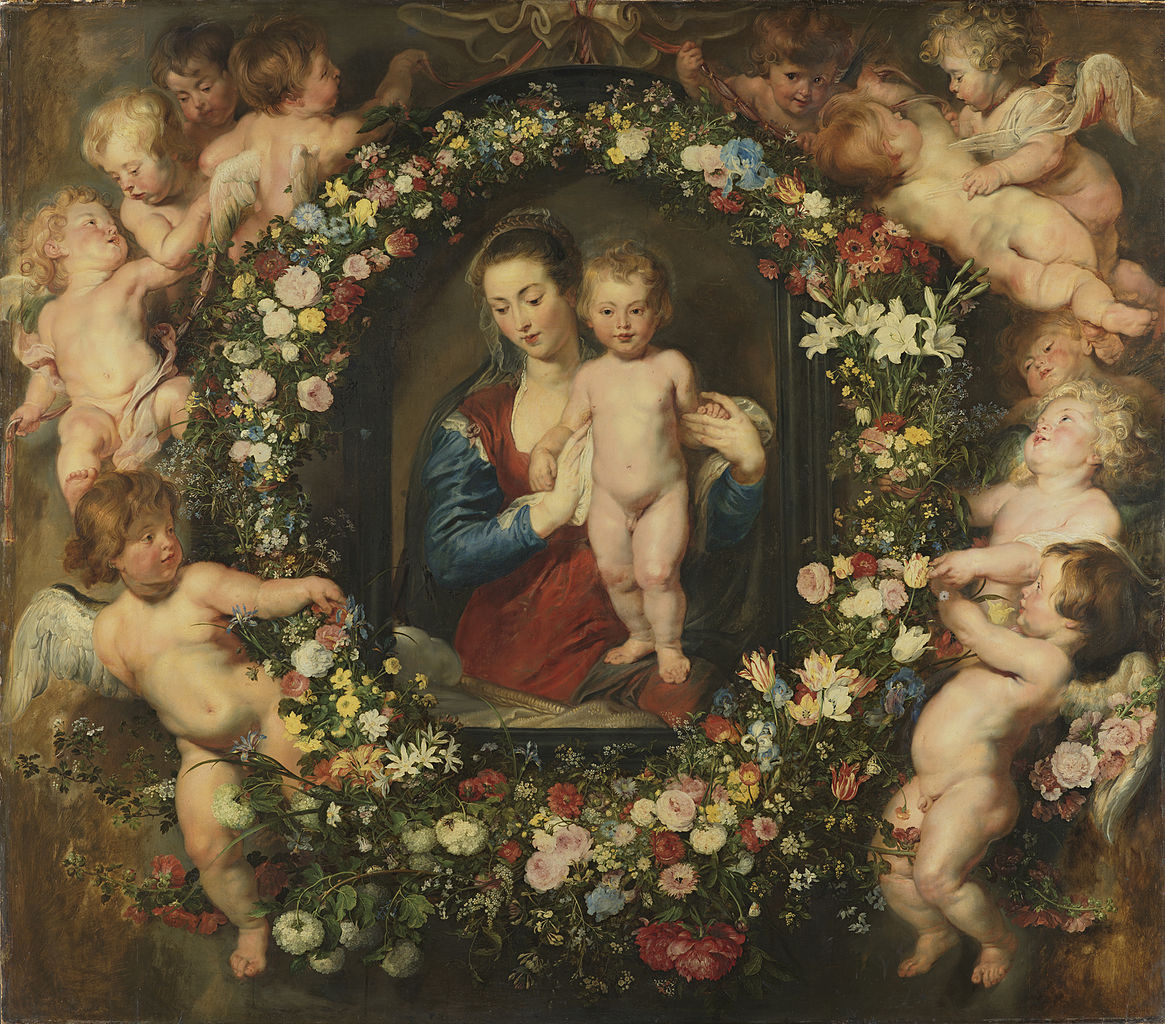}%
 \label{fig:collab_hist_a}%
}\hfill
\subfloat[Collaborative Collage]{%
 \includegraphics[height=3.55cm]{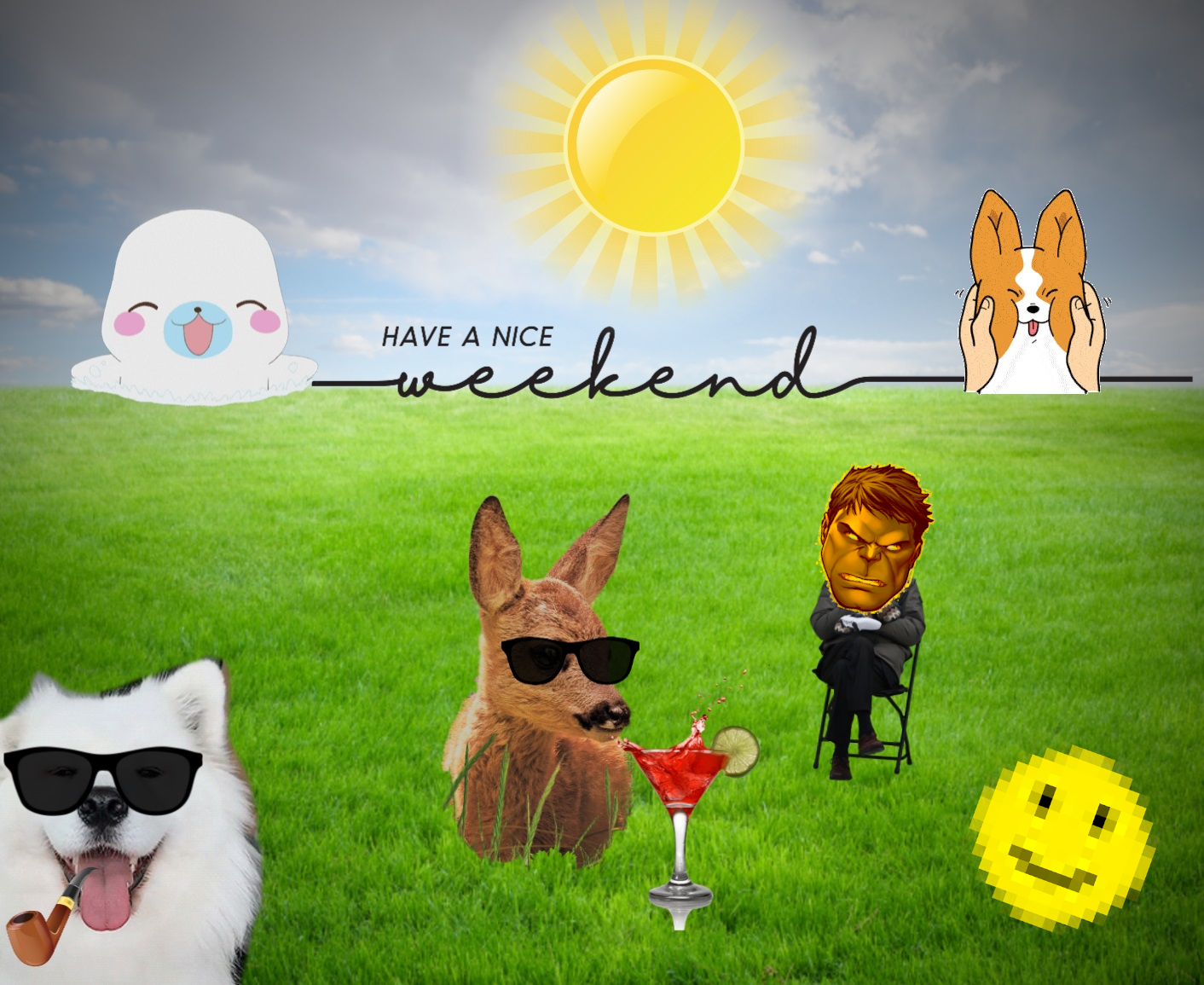}%
 \label{fig:collab_hist_b}%
}
\caption{\protect\subref{fig:collab_hist_a} An early example of collaboration between \emph{Jan Brueghel the Elder} and \emph{Peter Paul Rubens} approx.~1617. \protect\subref{fig:collab_hist_b} A collage created collaboratively using our web-based system. It comprises the blending of multiple layers, vector strokes, and image processing operations (\eg vignetting and pixelation).}%
\label{CART:Fig:CollaborationHistory}
\belowfig
\end{figure}

\noindent Collaboration between visual artists dates back to as early as late \nth{16} century (\Cref{fig:collab_hist_a}). 
In the modern era, this practice continued resulting in various masterpieces~\cite{Lee2017}. 
However, its adaptation in the digital domain is progressing only slowly (\Cref{fig:collab_hist_b}). 
Even though there exist collaborative applications mimicking a shared whiteboard -- allowing for doodling and/or simple manipulations of a shared image. 
A system for real-time collaborative editing of raster-based images at different levels of functionality or control -- similar to common image editing desktop applications (\eg Adobe Photoshop or GIMP) -- does not exist to the best of our knowledge~\cite{Isenberg2016}.

In general, multi-user systems where the actions of one user must quickly be propagated to the other collaborator are referred to as real-time groupware~\cite{Keith1997}. 
During recent years, various instances of such systems emerged into today's distributed, collaborative working environments. 
For example, systems such as Google Workspace or Microsoft Office 365 Online edition, various massively multiplayer online games, or NVIDIA's Omniverse platform for 3D contents. 
All such systems have in common that (1) a document instance or a shared context is hosted by a server(s), is (2) synchronized using a service, and (3) can be manipulated by multiple participants. 
The most relevant characterizing aspects of a real-time groupware system, according to Ellis and Gibs~\cite{Ellis1989}, are as follows:

\begin{itemize}
\item \textbf{Interactive and real-time (\Aspect{Aspect:Interactive}):} \ie response times must be short and notification times must be comparable to response times.
\item \textbf{Distributed (\Aspect{Aspect:Distributed}):} \ie in general, one cannot assume that the participants are all connected to the same machine or even to the same local area network.
\item \textbf{Volatile and Ad-hoc (\Aspect{Aspect:Adhoc}):} \ie participants are free to come and go during a session and generally are not following a pre-planned script. It is not possible to tell a priori what information will be accessed.
\item \textbf{Focused (\Aspect{Aspect:Focused}):} \ie during a session there is high degree of access conflict as participants work on/modify the same data.
\end{itemize}

\noindent A popular application that fulfills all of the above criteria is Google Docs. 
The cloud-based service provided by Google has revolutionized the way people edit documents collaboratively.
However, when editing text, most standard algorithms do not consider the complete structure of the document and make use of per-line diffing and merging.

\paragraph{Challenges}
The above approach cannot be extended for images in a straightforward manner: while \ref{Aspect:Interactive} and \ref{Aspect:Distributed} largely pose specific technical challenges (\eg undo/redo functionality and latencies), \ref{Aspect:Adhoc} and \ref{Aspect:Focused} reflects on the spatial, structural, and temporal features of collaborative raster-image editing.
Existing collaborative whiteboard applications maintain their state by tracking the brush strokes of individual clients. 
They allow users to doodle/sketch on top of the image but hardly provide any tools for image editing itself.
In particular, these are missing an integrated approach for the manipulation of the raster data using different image filtering operations.
On the other hand, existing web-based image editing tools are not collaborative. 
A system that is quite close to what we aspire is the Google Draw, a functionality provided as part of Google Workspace.  
Even though it allows users to collaboratively edit attributes of a shared image, the range of per-pixel edits is limited.

We adopt a human-centered design process to identify the challenges associated with a real-time collaborative image editing system. 
To this end, we design and conduct a preliminary user study with twenty-seven participants through a questionnaire. 
The answers to the above questionnaire identify key design principles mainly focused on communication and synchronization.
We prototype and iterate on the design of our collaborative system. 
To understand whether our strategies achieve the design goals, we perform a post-deployment user study with six different groups (two or three persons each).
Our system was successful to a large extent.
Moreover, participant’s experiences and perspectives offer further guidance for improvement.

\paragraph{Approach \& Contributions}
We aim to create a web-based collaborative image editing application that provides a wide range of edits. 
To this end our contributions are: 

\begin{enumerate}
\item A web-based application that allows multiple users to collaboratively edit images, while satisfying \ref{Aspect:Interactive} to \ref{Aspect:Focused} properties. 
The Web-App consists of a responsive \ac{GUI} which makes it possible for users to access the application via a smartphone or a tablet. 
\item A browser-based rendering framework that enables a wide range of image manipulations along with sketching/doodling functionality.
\item Results of our preliminary and post-deployment user studies that identify key design principles for a real-time collaborative image editing system.  
\end{enumerate}

\noindent For it, we choose the following approach.
\Cref{CART:Sec:RelatedWork} reviews and analyzes related work and existing tools on collaborative editing of graphics.
Based on these, a preliminary user study on the current state of real-time collaborative image editing and associated tasks is conducted (\Cref{CART:Sec:PreliminaryUserStudy}).
\Cref{CART:Sec:SysOverview} describes a system overview of our prototypical implementation of basic server and client functionality.
\Cref{CART:Sec:Evaluation} evaluates the implementation through a post-deployment user study.
We summarize our findings and potentials for future work and research in \Cref{CART:Sec:SystemOverview}.

\begin{table*}[tb]
	\centering%
	\caption{Comparison of various web applications for the editing of raster images with respect to different features.}
	\label{CART:Tab:Comparison}
	\scriptsize%
	\begin{tabular}{|l|c|c|c|l|c|c|l|}
		\hline
		\textbf{Application} & \textbf{Layer} & \textbf{Direct Manip.} & \textbf{Undo/Redo} & \textbf{Image Filtering} & \textbf{Data Type} & \textbf{Resp. \acs{GUI}} & \textbf{Collaboration Type} \\
		\hline
		\href{canvaspaint.org}{canvaspaint.org} & No & Yes & Yes &None & Raster & Yes & None\\
		\href{pixlr.com}{pixlr.com}& Yes & Yes & Yes &Destructive & Raster & No & None\\
		\href{photopea.com}{photopea.com} & Yes & Yes & Yes &Destructive & Raster & No & None \\
		\href{draw.chat}{draw.chat}& No &Yes  & Yes & None & Vector & No &  Synchronous \\
		\href{aggie.io}{aggie.io} & Yes & Yes & Yes &None & Raster & Yes & Synchronous\\
		Google Draw & No & No &  No &Non-destructive & Raster & Yes & Synchronous\\
		Adobe Creative Suite & Yes & Yes & Yes &Destructive & Both & Yes & Asynchronous \\
		\hline
	\end{tabular}
	\belowfig
\end{table*}

\section{Background \& Related Work}
\label{CART:Sec:RelatedWork}

\noindent The challenges associated with collaborative image editing has two aspects: the conceptual/design level and implementation level, which nowadays demands web-based approaches using services. 
The existing web-based applications mainly address sketching and/or designing functionalities.

\paragraph{Collaborative Graphics Editing}

\noindent One of the earliest study towards the desired characteristics of a collaborative graphics editing system was performed by Sun and Chen~\cite{Sun2002b}. 
They propose a formal specification for conflict resolution, versioning, and consistency maintenance for such systems.
Design analysis of visual analytic tools has been explored by Heer and Agrawala~\cite{Heer2008}, where they propose techniques to improve shared context and awareness, and provide suggestions to increase engagement.
As a specific instance, Salvati \etal~\cite{Salvati2015} and Calabrese \etal~\cite{Calabrese2016} analyze collaborative mesh manipulation by robustly sharing and merging version histories in real-time.
In recent work, Gao \etal~\cite{Gao2018} map the two-dimensional drawing area into the linear structure and correspondingly transform the two-dimensional graphical operations to linear operations for collaborative editing. 
In order to prevent consistency conflicts, Wu \etal~\cite{Wu2019} propose the \ac{CGCE} algorithm.
Both Gao \etal and Wu \etal implement their solution using the latest web-based technologies, however, their system only allows for sketching or primitive geometric figure manipulation.
For the purpose of cooperative image editing, Zhai \etal~\cite{Zhai2005} develop a method using  wireless communication over mobile phones. 
Nevertheless, they only consider simple atomic operations of \emph{import, export, update}, and \emph{commit}. 
Novakova \etal~\cite{Novakova2013} developed a tool specifically for collaborative sketching, suitable for architectural communication.
In comparison, our layer-based rendering framework can handle a variety of image edits and also provides brushing and sketching functionality. 

\paragraph{Web-based Sketching and Designing}

\noindent Web-based collaborative whiteboards allow users to ideate and collaborate visually, \eg Aggie.io or Draw.chat. 
In comparison, collaborative design tools are more recent and focused on creating new designs by arranging images as multiple layers, \eg Figma, Canva, or AdobeXD.  
However, both types of applications have hardly any image editing functionality. 
The existing web-based image editing applications can be used for per-pixel processing, but are not collaborative in nature, \eg Photopea or Pixlr. 
In a very recent development, Adobe now allows for asynchronous collaboration for raster and vector images. 
Nonetheless, our goal is to provide a real-time synchronous collaborative environment. 
\Cref{CART:Tab:Comparison} compare existing web-based photo-editing and whiteboard applications regarding the following aspects: 

\begin{itemize}
\item \textbf{Layer (Yes/No):} The application does support layering of multiple images. This allows for an increased function scope and assumes a complex data model.
\item \textbf{Direct Manipulation (Yes/No):} The application does support direct manipulation of image contents, \eg using brushing or transform functionality.
\item \textbf{Undo/Redo (Yes/No):} The support of undo/redo functionality facilitates error-tolerance while using direct manipulation metaphors.
\item \textbf{Image Filtering (None/Destructive/Non-destructive):} An application supports the usage of single or multiple destructive/non-destructive image filtering operations.
\item \textbf{Data Type (Raster/Vector/Both):} The application can handle raster, vector, or both types of input data.
\item \textbf{Responsiveness (Yes/No):} The \ac{GUI} of the application support responsive layout of components, thus supports desktop and mobile devices with varying screen sizes.
\item \textbf{Collaboration Type (None/Synchronous/Asynchronous):} A collaborative application enables multiple clients to modify image data simultaneously. This requires communication between clients and modeling of messages reflecting the editing process.
\end{itemize}

\noindent Our web-based collaborative system provides sketching and designing functionality along with image manipulations.
Moreover, we enable synchronous collaboration among users.
It provides all features compared in \Cref{CART:Tab:Comparison} while operating on layered raster images.

\section{Analysis and Preliminary Considerations}
\label{CART:Sec:PreliminaryUserStudy}

\noindent This section reports on a preliminary user study~(\Cref{CART:SubSec:PreliminaryUserStudy}), conducted to analyze the major requirements for real-time collaborative image editing and elaborate potential conflicts to be addressed in collaborative editing~(\Cref{CART:SubSec:Conflicts}).
\subsection{Preliminary User Study}
\label{CART:SubSec:PreliminaryUserStudy}

\noindent To better understand the design requirements of a real-time collaborative image-editing application, we designed and conducted a preliminary user study using a questionnaire. 
Subsequently, we analyzed results and major findings, identified main use cases, and created \ac{GUI} design sketches.

\paragraph{Participants and Study Design}

We selected participants who have experience in collaborative image editing with existing technologies.
They belong to a broad range of background and have performed image editing: for casual creativity and/or as a professional activity. 
A total of \num{27} participants answered the \num{17} questions. 
The questions broadly addressed the following aspects: ($i$)~\emph{How do you perform collaborative image editing tasks?} and ($ii$)~\emph{What are the challenges associated with it?}. 
The challenges thus identified are used as the basis for designing our system. 

\paragraph{Summary on Challenges in Collaboration}

The foundation of any collaborative task is efficient \emph{communication}, which also reflects in our survey answers:
\enquote{\emph{Communication is everything, it is sometimes hard to get an artistic idea thru}} (P7).
\enquote{\emph{Communicating who edits what and how}} (P9).
The lack of communication is not only restricted to high-level requirements and task sharing but also low-level details such as data/edits synchronization:
\enquote{\emph{Staying in sync, keeping a history of changes, knowing what the partners already have done}} (P5).
\enquote{\emph{Handling data conflicts, know on which parts or region my teammates are currently working on, handling different versions...}} (P10).
To mitigate the above problems, users make use of existing communication channels.
However, such an approach seems to be quite inefficient in terms of both data and time:
\enquote{\emph{It takes a lot of time sending images back/forward and see when progress is made}} (P11).
\enquote{\emph{Sharing huge files of raw pictures with the team and keeping them in sync}} (P6).

\paragraph{Design Inference}

Concerning the above challenges, we choose an integrated messaging functionality for our system. 
Our WebGL-based rendering framework allows for image edits that are visible in real-time among the collaborators, further enabling low-level communication.
The user edits are maintained as part of session management providing for a consistent editing environment.
Data conflicts are handled with complementary update processes on server and client-side (\Cref{CART:SubSec:Server}).
Similar to the variety of challenges and its coping strategy, there is a range of application scenarios where collaborative image editing can be used:
\enquote{\emph{logo creation}}, \enquote{\emph{poster designing}}, \enquote{\emph{creative editing}}, \etc.
We support such varying application scenarios by providing \ac{VCA}-based image editing along with a mouse/pen/touch-based sketching interface. 

\subsection{Potential Conflicts in Collaborative Editing}
\label{CART:SubSec:Conflicts}

\noindent There are several potential conflicts arising in real-time collaborative image editing systems, especially if these support a variety of tools being applied to multiple layers.
Specific to our approach this concerns challenges arising from ($i$) limited attention-bandwidth and ($ii$) synchronous as well as ($iii$) asynchronous editing conflicts.
Considering users operating in the same sessions, our system offers tools to approach the above challenges.

\paragraph{Limited Attention-bandwidth}

While performing editing tasks, such as painting or designing, a user focuses on the immediate effect of the current tool.
Thus, the user has a limited attention-bandwidth and is usually not aware of the changes performed by other users in the document.
For example, adding new layers impacts the layer order and often interrupts the user's workflow, and can quickly lead to confusion.
To counterbalance this, a document version history is offered that enables users to comprehend the performed changes over time.

\paragraph{Synchronous Editing Conflicts}

There are various causes for conflicts in synchronous data editing, \eg users use the same/different tool on a given layer or a layer is about to be removed that is used by others users.
Instead of making tools modal, we choose to raise awareness by indicating that another user is using the same tool or has selected the same data using visual feedback.
For this purpose, colored hints (lines) visualize which user(s) currently select a layer and tool respectively (\Cref{CART:SubSubSec:UserSpecificVisualFeedback}).
Moreover, the cursors of all users are depicted in the respective avatar color (\Cref{CART:SubSubSec:SessionHandling}).

\paragraph{Asynchronous Editing Conflicts}

These conflicts are often caused if several users simultaneously edit the same layer or due to interruptions of unfinished tasks, \eg a user is interrupted in its current workflow but wants to continue his/her work later on.
This can cause conflicts if other users are not aware of this state and meanwhile perform a task on the same data.
To approach this, we introduced an \emph{exclusive-lock} functionality for a layer, \ie a user can forbid editing of a layer for everyone except himself (\Cref{CART:SubSubSec:BasicEditingFeatures}).
To avoid deadlock scenarios, a layer can be \emph{exclusive-unlocked} by others users.
In this case, the user who initiates the exclusive-lock is notified accordingly.

\begin{figure}[tb]
	\centering
	\includegraphics[width=1.0\linewidth]{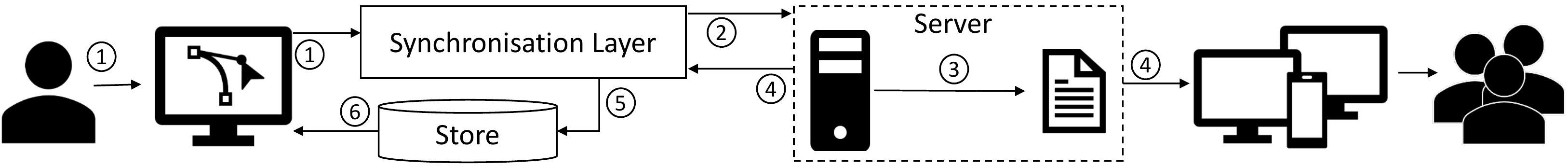}
	\caption{Sequence of client-server communication: 
		(1) user modifies project,
		(2) modified parameters are processed in synchronization layer,
		(3) a change request is sent to server,
		(4) server updates document
		(5) if successful, updates are sent to all clients,
		(6) synchronization layer updates the local store,
		(7) changes are applied in the \ac{GUI}.}
	\label{CART:Fig:Communication}
	\belowfig
\end{figure}
 
\begin{figure}[tb]
\centering
\includegraphics[width=1.0\linewidth]{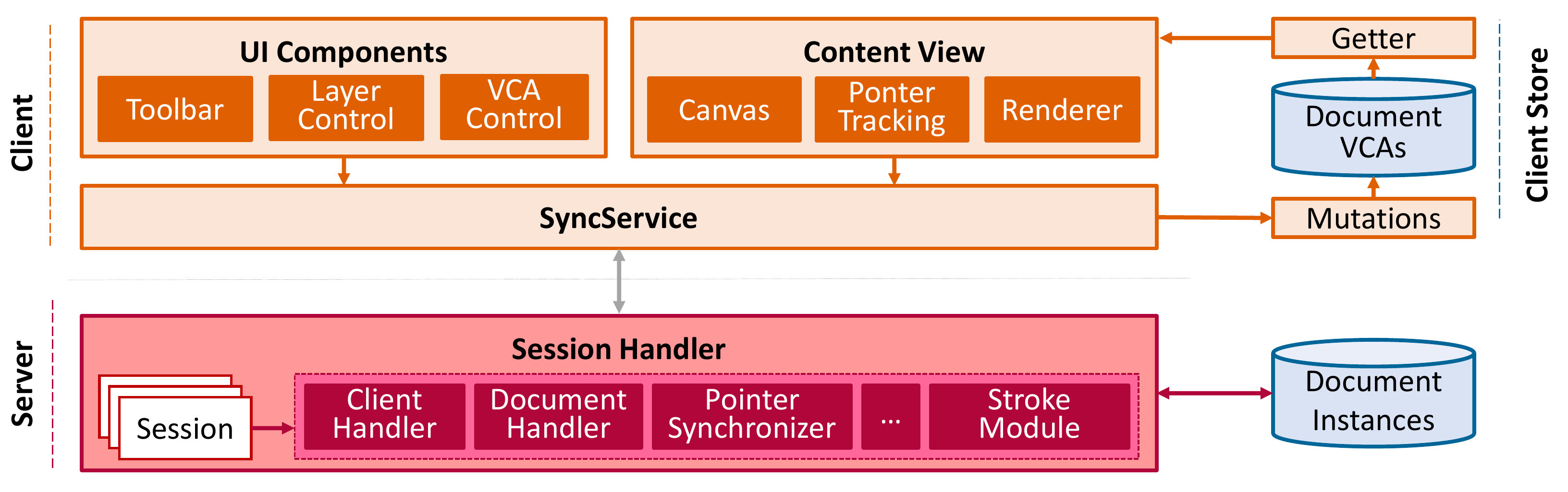}
\caption{Overview of the client-server architecture used for our system.
The server synchronizes and propagates project modifications among clients through different session handlers.
A user can modify the project through the \ac{GUI} components and the content view.
The cyclic update process on the client-side prevents version conflicts.}
\label{CART:Fig:Architecture}
\belowfig
\end{figure}

\section{System Overview}
\label{CART:Sec:SysOverview}

\noindent We develop our system as a \ac{SPA} that can be used on desktop systems and on mobile devices.
It enables sketching, image adjustments, and creation of image collages among multiple clients in real-time.
An overview of our system with a depiction of a client to server communication and vice versa is presented in \Cref{CART:Fig:Communication}. 
To achieve this distributed structure of a real-time groupware system, we develop an extensible client-server architecture~(\Cref{CART:Fig:Architecture}).
The server is mainly responsible for session handling and synchronization, and maintaining communication among clients~(\Cref{CART:SubSec:Server}).
The client transmits and consumes messages~(\Cref{CART:SubSec:Communication}) which represent editing actions and perform client-side rendering~(\Cref{CART:SubSec:Client}).   

\subsection{Server Components and Functionality}
\label{CART:SubSec:Server}

\noindent The main task of the server component is to maintain the session documents, manage and provision its state, and handle the communication between the clients.

\subsubsection{Session Document}

\setlength{\intextsep}{10pt}%
\setlength{\columnsep}{10pt}%
\begin{wrapfigure}[10]{R}{0.5\linewidth}
\vspace{-0.75\baselineskip}
\centering
\includegraphics[width=1.0\linewidth]{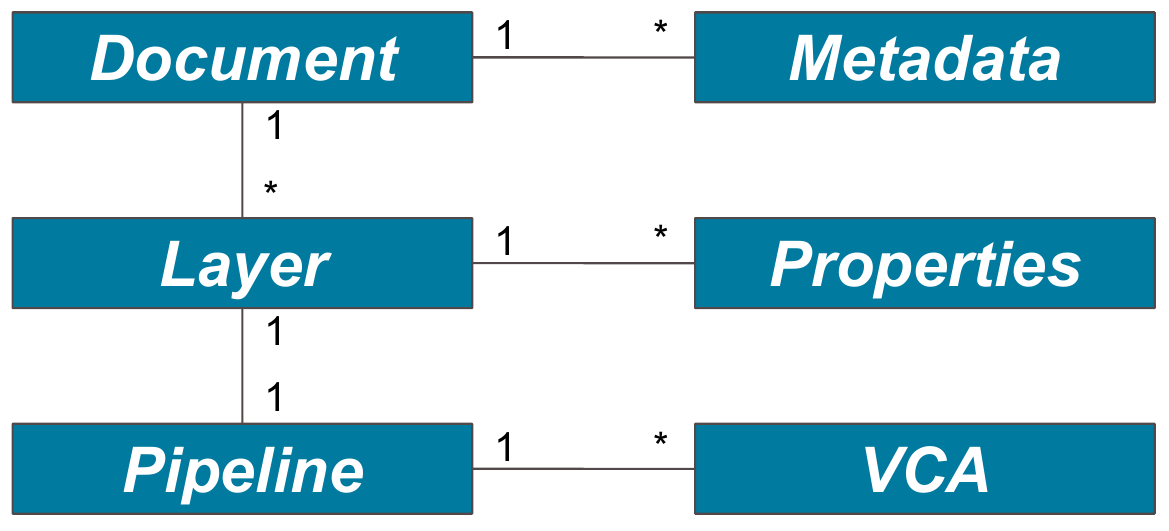}%
\caption{High-level structure of a session document comprising multiple layers with multiple \acp{VCA}.}
\label{CART:Fig:DocumentStructure}%
\end{wrapfigure}

The document structure (\Cref{CART:Fig:DocumentStructure}) is inspired by the OpenRaster file format. 
A document consists of metadata about the project, \eg creation date, version, resolution, \etc. 
All used images in a project are stored as layers in an array in the document. 
These layers contain property information about the image, \eg transformations, visibility, \etc. 
Images can be edited using pipelines, \ie \acp{VCA} are applied consecutively.
The effect name of each used \ac{VCA} and its parameters are stored in an array (pipeline) in the layer object. 
This enables several customization possibilities on a per-layer level and per-\ac{VCA} level.

\subsubsection{Session Handling}
\label{CART:SubSubSec:SessionHandling}

\setlength{\intextsep}{10pt}%
\setlength{\columnsep}{10pt}%
\begin{wrapfigure}[12]{R}{0.45\linewidth}
\vspace{-0.85\baselineskip}
\begin{lstlisting}[language=json,firstnumber=1,abovecaptionskip=5pt,captionpos=b,caption={Exemplary message structure of a \texttt{newPath} action in the \texttt{drawing} module.}, label={CART:Lst:MessageStructure}]
{"module": "drawing",
 "message": {
  "newPath": {
   "timeStamp": "1617804631471", 
   "clientId":  "m82pY9bvAeIAAAH", 
   "color":     "#795EB3", 
   "width":     "10", 
   "path": [["M",446.99,38],
            ["Q",447,38,448,38],
            ["Q",449,38,449.5,38],
            ["Q",450,38,451,38.5],
            ["Q",452,39,452.5,39],
            ["L",453.01,39]]}}}
\end{lstlisting}
\end{wrapfigure}

At the initial state, all stored documents are read and each one is assigned a unique identifier. 
These sessions contain different handlers for broadcasting client information, \eg pointer positions and actions, and/or updating the document and notifying clients.
When a client connects to the server, the server responds with a session overview.
After registering for a session, the user's socket is subscribed to all handlers of the specific session on the server-side. 
By registering to the server, the client receives its unique server-socket \ac{ID} that is stored in the local storage of the browser. 
If the client disconnects, it re-sends its assigned~\ac{ID} when reconnecting to the server and thus is recognized again.
Moreover, a unique color is assigned to the client, which also serves as the default brush color.

Since several users can work simultaneously in one session, we have a high degree of access-conflict. 
The server treats incoming changes as \enquote{first come, first serve} and, hence, defines the order of updates which is then sent to all subscribed clients.
The main logical conflicts are resolved at server-side, \eg if a user deletes a given layer while another user edits it, the latter change request is dropped.
Remaining access-conflicts, which are not mutually exclusive, are then handled at the client-side, \ie the last executed update will define the modified session state.
Thus, session-handling is important to maintain synchronization among clients, a key requirement (\Cref{CART:Sec:PreliminaryUserStudy}) for such a system.

\begin{figure*}[!tp]
	\centering
	\subfloat[Editing Tools and Views]{%
		\includegraphics[width=0.5\linewidth]{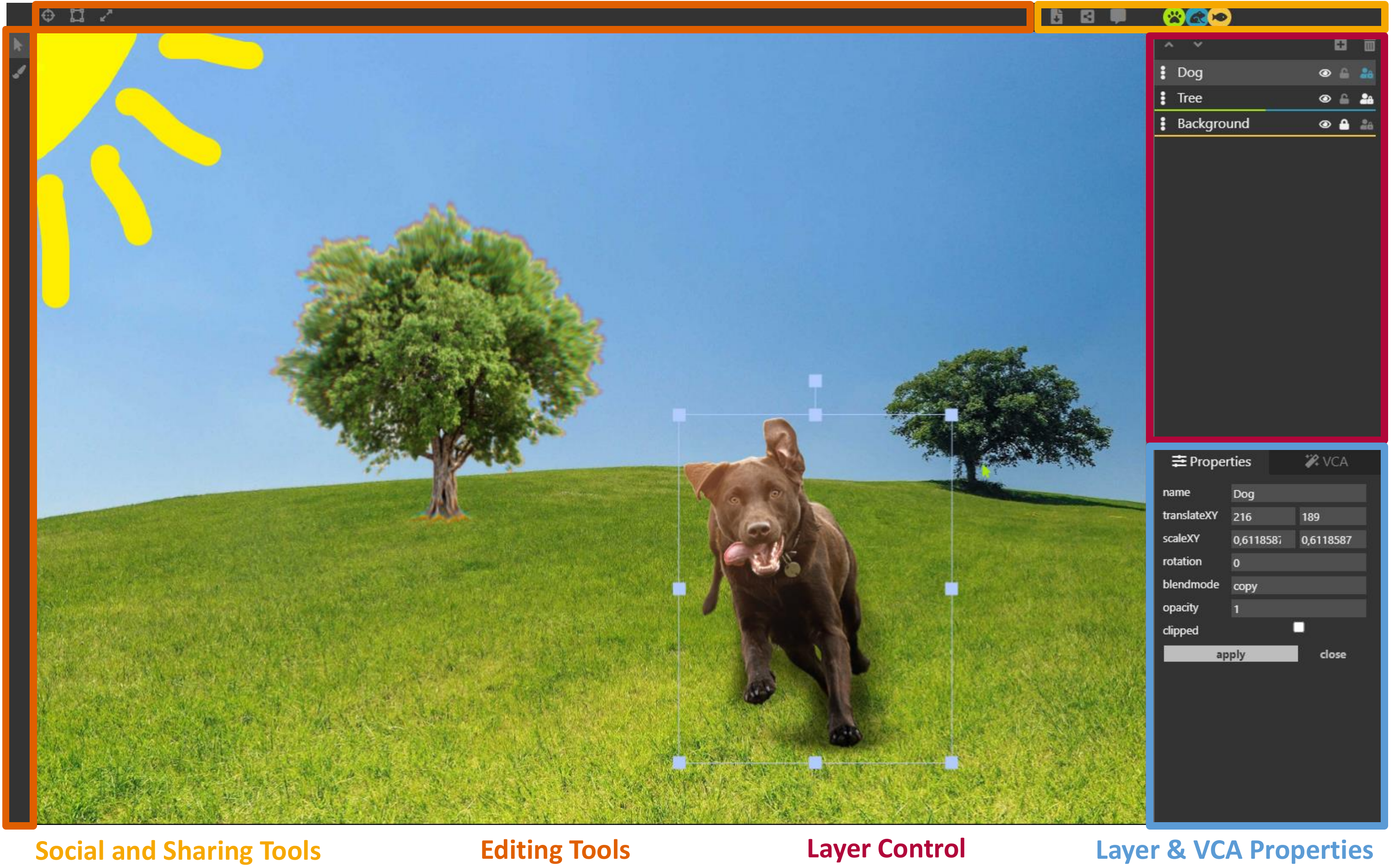}%
		\label{CART:Fig:UI:a}%
	}\hfill
	\subfloat[User-specific Visual Feedback]{%
		\includegraphics[width=0.5\linewidth]{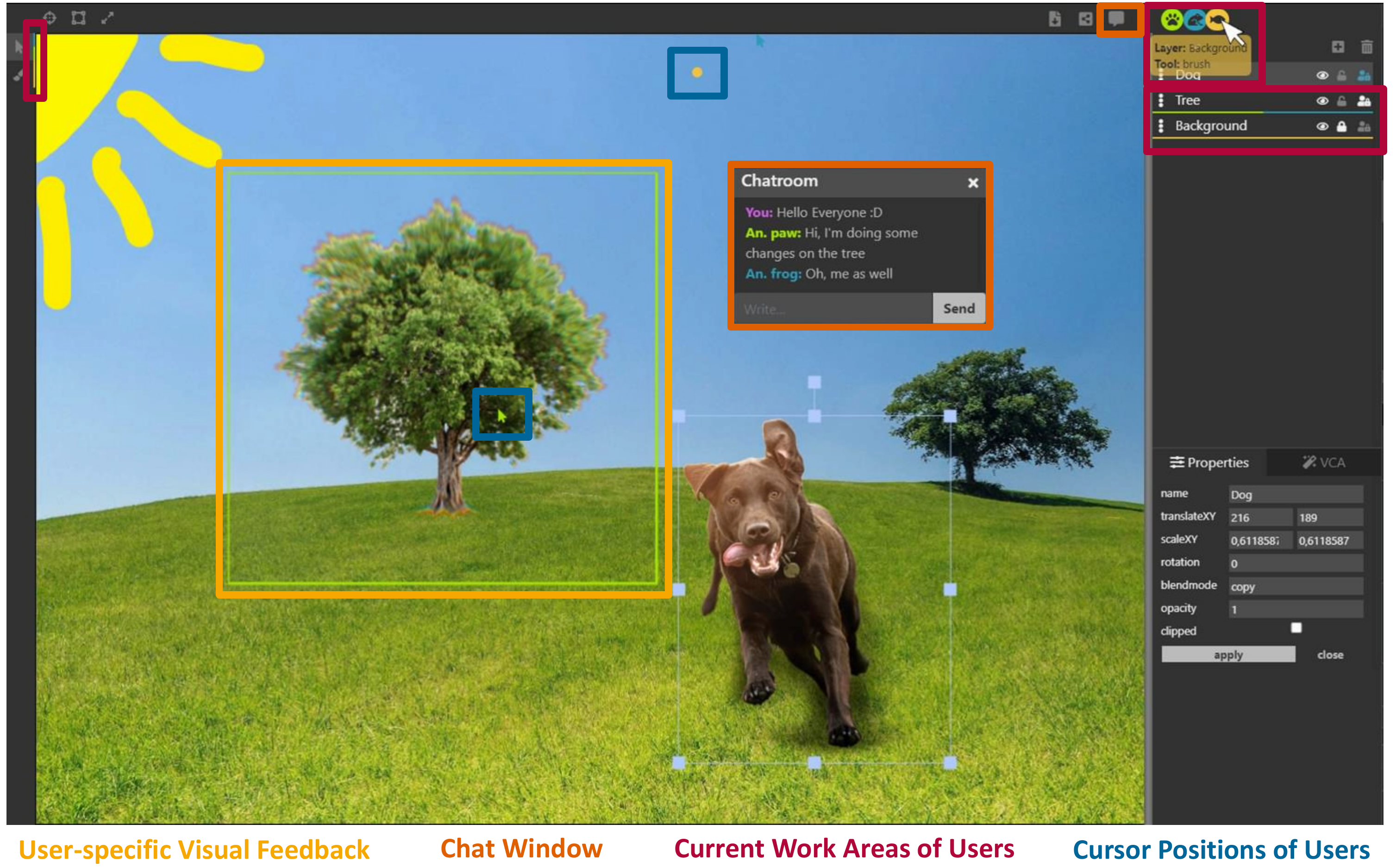}%
		\label{CART:Fig:UI:b}%
	}
	\caption{Our \ac{GUI} provides a variety of \protect\subref{CART:Fig:UI:a} editing tools and \protect\subref{CART:Fig:UI:b} user specific visual feedback to visually communicate the tool and objects currently operated by other users in order to mitigate the risk for potential editing conflicts.}%
	\label{CART:Fig:user_interface}
\end{figure*}

\subsection{Protocol for Client-Server Communication}
\label{CART:SubSec:Communication}

\noindent For the communication between the server and multiple clients, we design a simple protocol that suffices the following requirements:
($i$) it has a simple yet extensible message structure to facilitate easy development and allows the integration of future features;
($ii$) it is suitable for fast message (de)serialization to reduce the run-time overhead for clients and server.
The clients employ a WebSocket connection to send events to the server, which are then broadcasted to the remaining clients.
Both client and server listen for events and process the incoming data accordingly.
The sent data includes information about the applied project changes as well as other aspects such as timestamp and client \ac{ID} that allow for change-history maintenance and enable change traceability among the users.
An exemplary message structure (based on \ac{JSON} standard format) for a \texttt{newPath} event is depicted in \Cref{CART:Lst:MessageStructure}.
The above allows for efficient communication among clients, which is a necessity  (\Cref{CART:SubSec:PreliminaryUserStudy}) for our system.

\subsection{Client Components and Functionality}
\label{CART:SubSec:Client}

\noindent The rendering of the raster images is performed entirely on the client-side using WebGL 2.0.
The front-end is developed using Vue and the Vuex framework is employed for global storage. 
Moreover, we make use of Fabric.js to facilitate layer control for canvas rendering. 

\subsubsection{Update Logic}

Since all users can potentially work on the session document simultaneously, resulting data-conflicts are required to be handled properly. 
For it, we propose the communication process among clients as depicted in~\Cref{CART:Fig:Communication}.
After the user changes a parameter, a change-request is sent to the server over a synchronization service. 
The above service also listens to all change-events from the server. 
If the request is accepted by the server, it will notify all clients. 
The service then modifies the store and the \ac{GUI} is updated accordingly. 
The user is not allowed to update the local store directly to prevent version conflicts.
A small update delay is barely noticeable because of quick socket communication.
In case of a parameter-update conflict, when two users update the same value simultaneously, the last request processed by the server is considered the final version.
However, only one user at a time should be able to directly manipulate or transform a layer.
For it, during such operations, the layer will be implicitly locked w.r.t. its transformation properties.

\begin{figure}[t]
	\centering
	\includegraphics[width=1.0\linewidth]{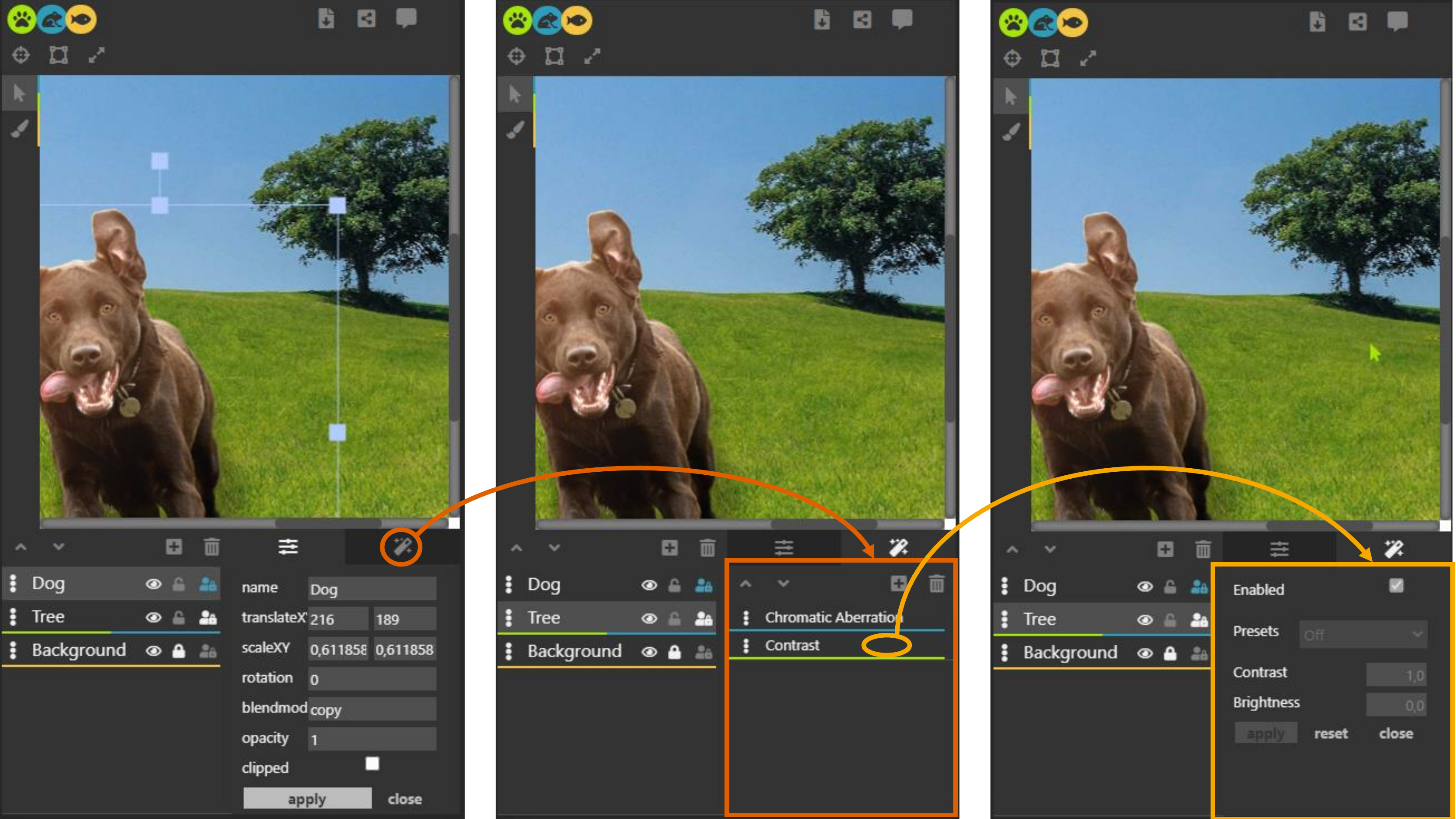}
	\caption{The responsive \ac{GUI} layout hides editing components if the screen size is too small. The functionality can be easily expanded by pressing the respective button.}
	\label{CART:Fig:responsive_ui}
	\belowfig	
\end{figure}

\subsubsection{GUI Structure/Schematics}
We assume that the target audience is familiar with some raster-image editing software and therefore decided to re-use \ac{GUI} concepts from common image-editing applications.
Thus, tools such as brush and selection are located in a vertical icon toolbar on the left with additional control parameters on an upper horizontal bar~(\Cref{CART:Fig:user_interface}).
The object property panels (\eg of layers or \acp{VCA}) are both located on the right side of the raster graphic.
A user can directly interact with the canvas by drawing on a layer or transforming it.
Since the image takes up most of the available space for direct editing, the remaining \ac{GUI} components are arranged compactly with informative icons to ensure intuitive usability. 
Most applications for raster image editing do not support a responsive design. 
For smaller screen sizes, \eg mobile or tablet, this is problematic, because many operations must be clearly represented with large buttons for easy access.
Therefore, we hide certain components, which are displayed via a responsive layout if required~(\Cref{CART:Fig:responsive_ui}).
Since the components themselves do not differ between screen sizes, the user can easily switch between desktop and mobile devices without adapting to a new \ac{GUI}.
The generic project tool buttons for downloading the final image \inlineimage{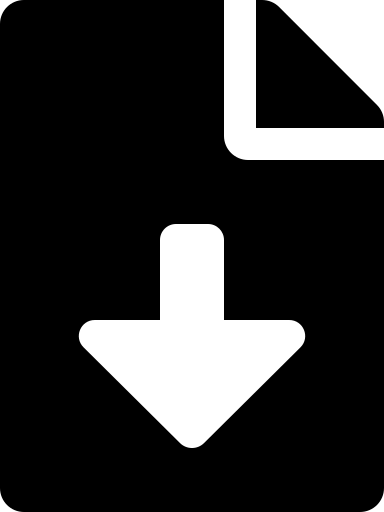}, sharing the project \inlineimage{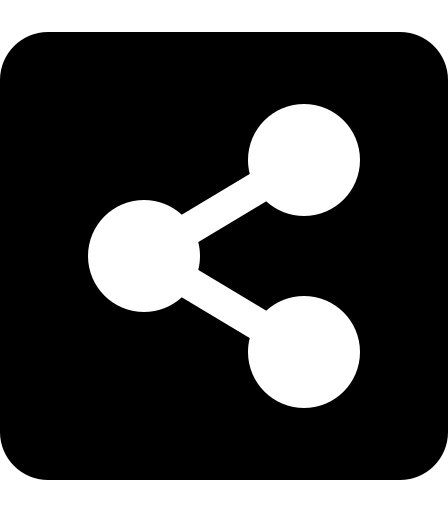},
or messaging other users working on the project \inlineimage{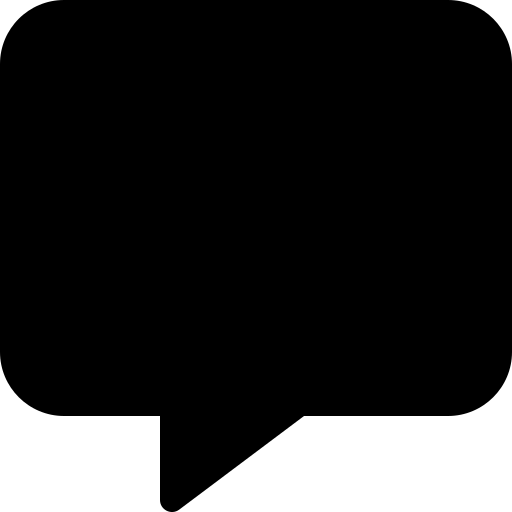}, are placed on top of the editing components.

\subsubsection{User-specific Visual Feedback}
\label{CART:SubSubSec:UserSpecificVisualFeedback}
For a coordinated workflow among the clients, the respective selected layer, VCA and the tool of each user is highlighted~(\Cref{CART:Fig:UI:b}).
This allows a user to reproduce canvas changes made by another user.
Moreover, this potentially avoids editing conflicts or parameter overwrites as the user can see if someone else has selected the same layer or \ac{VCA}.
Similar to other collaborative web-apps, an overview of currently active users is depicted in the upper right corner.
On hovering over the user's icon, the respective username is displayed.
We can also get an overview of the user's working area by clicking on the user icon.
The cursor position on the canvas is broadcasted to the remaining clients and is displayed with the client's unique color identifier, assigned by the server.
The displayed cursor depends on the selected tool, \eg a pointer or a brush.
Additionally, if a user selects or transforms a layer, it is highlighted with the respective client's color.
This way, all participants obtain an overview of the active objects of other users.

\subsubsection{Basic Editing Features} 
\label{CART:SubSubSec:BasicEditingFeatures}
Within a project, layers can be added, deleted, and reordered with simple button clicks in the layer control panel. 
For each layer, visibility can be set \inlineimage{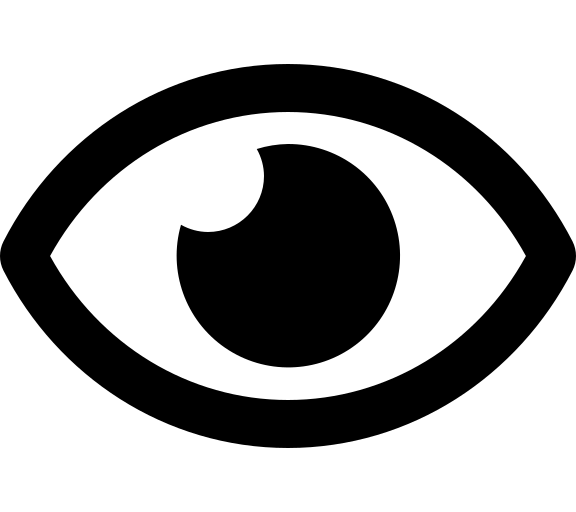} and the layer itself can be locked/unlocked \inlineimage{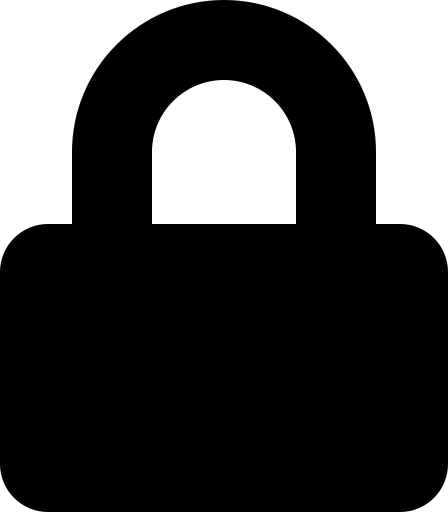}.
To further enable collaboration we introduce an \emph{exclusive-lock} button \inlineimage{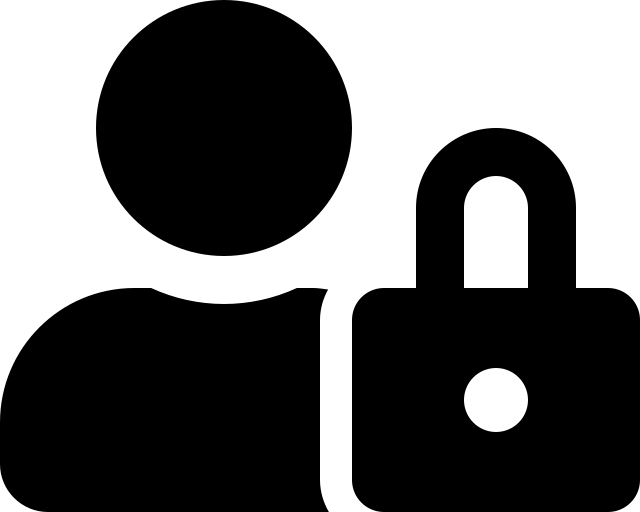}.
Analogous to the lock functionality, a user can disable a layer via this button.
However, the layer will be locked for everyone except this user.
Other users can see who exclusively locked a layer and when.
By unlocking this layer, the original user gets a notification.
This way, a user can personally lock a layer and signal that he/she does not want interference from other users.
Depending on whether the user himself/herself exclusively locked the layer or not, this button is highlighted in a different color.
Thus, the user also has a visual overview of which layers he/she is currently working on.
Furthermore, for each layer, additional information is displayed in the panel below.
A user can switch between the different control settings through tabs, \eg the layer properties or \ac{VCA}.
Thus, this panel can be easily extended with additional editing features later on by adding new tabs, \eg viewing the respective layer version history or comments.
The main layer properties, \eg scale, rotation, opacity, are located in the properties tab.
In the \ac{VCA} tab, the user can add, delete, and reorder \acp{VCA} in the pipeline of the layer.
Each \ac{VCA} is adjustable and can be enabled or disabled.
All changes are applied to the image in real-time.
Depending on the selected tool in the left toolbar, the user has different options to manipulate layers.
The corresponding horizontal bar above it is generic and can thus be extended with further tools and settings.
So far, the following tools are available:
\begin{itemize}
	\item \textbf{Select:} A layer can be selected and transformed with the respective handles.
	Additional buttons for resizing and centering facilitate the use.
	\item \textbf{Brush:} One can draw with customizable brush size and color on the selected layer.
	Performed brush strokes can be undone/redone using the respective tool buttons.
\end{itemize}

\section{Post-deployment User Study}
\label{CART:Sec:Evaluation}



\noindent Additional requirements on functionality and user experience are often identified after a prototype is deployed and users have had a chance to try the software and provide feedback.
This valuable feedback will be used to improve the future iterations of our prototype. 
For the post-deployment study, we focused on the following three aspects: 
($i$) do users understand the general structure of the \ac{GUI},
($ii$) do users understand the visualization metaphors to avoid editing conflicts,
and ($iii$) are users satisfied with the prototype.

\subsection{Participants \& Apparatus}

\noindent We recruited \num{16} volunteers (\num{8} male, \num{8} female) in \num{6} different groups. 
The above participants use our system for the first time and were not part of the preliminary user study to avoid any inherent bias.
Each group had a variable number of participants between \numrange{2}{3} and volunteers were aged between the ages \num{21} and \num{34}. 
While all of them had experience with computers, \num{5} had no or only little experience with image editing applications. 
All of them had normal or corrected-to-normal vision and no known visual impairments. 
All the participants (except for one, who used an iPad) accessed our \ac{SPA} on a desktop/laptop system with a single monitor using standard web-browsers (Google Chrome: \num{7}, Mozilla Firefox:~\num{5}, Apple Safari: \num{2}, Microsoft Edge: \num{2}) and a computer mouse (two participants used trackpads). 

We conducted a supervised/observed study in remote sessions, each with a group of participants.
We were connected with them via an online Zoom meeting as they were guided and monitored at the same time. 
Each session had a length of approx. \SI{60}{\min} having the following structure.
First, each group received a brief introduction into the \ac{GUI} covering only editing tools as well as layer and \ac{VCA} controls (\SI{5}{\min}). 
Following this, each group is asked to collaboratively solve three tasks in sequence. 

\begin{figure}[tb]
	\centering
	\subfloat[Task-1 Input]{%
		\includegraphics[width=0.31\linewidth]{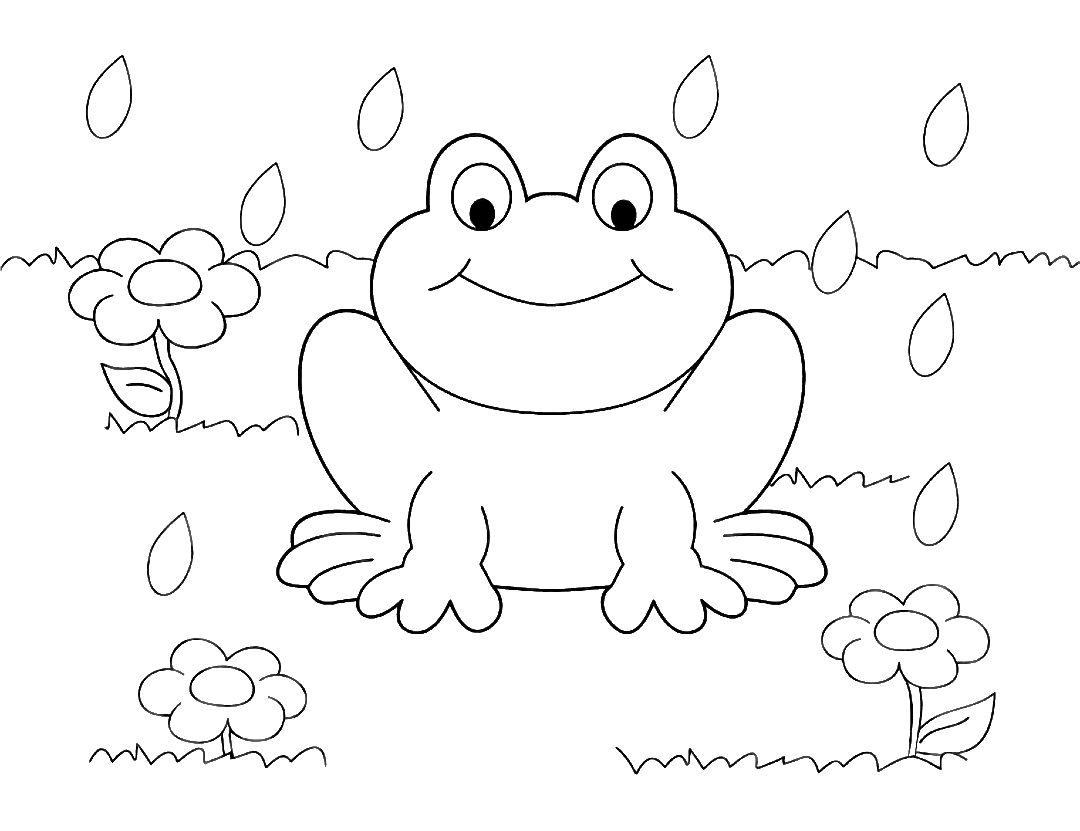}%
		\label{CART:Fig:task_1_in}%
	}\hfill
	\subfloat[Task-2 Input]{%
		\includegraphics[width=0.31\linewidth]{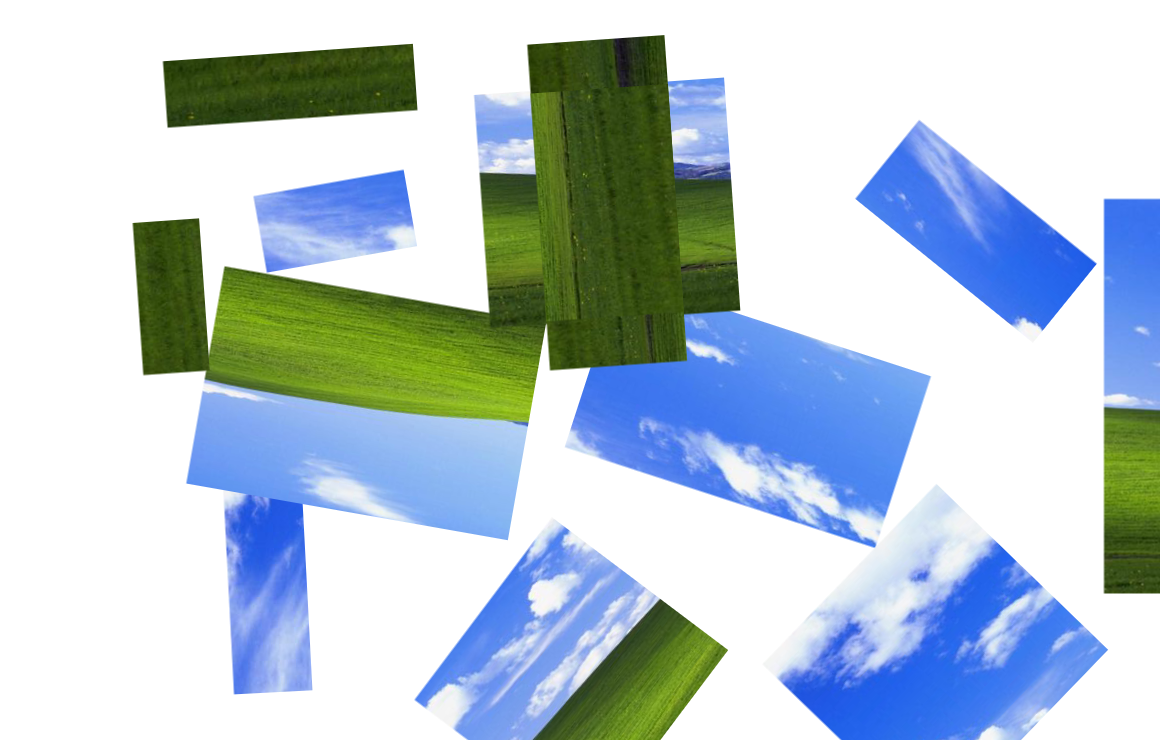}%
		\label{CART:Fig:task_2_in}%
	}\hfill
	\subfloat[Task-3 Input]{%
		\includegraphics[width=0.31\linewidth]{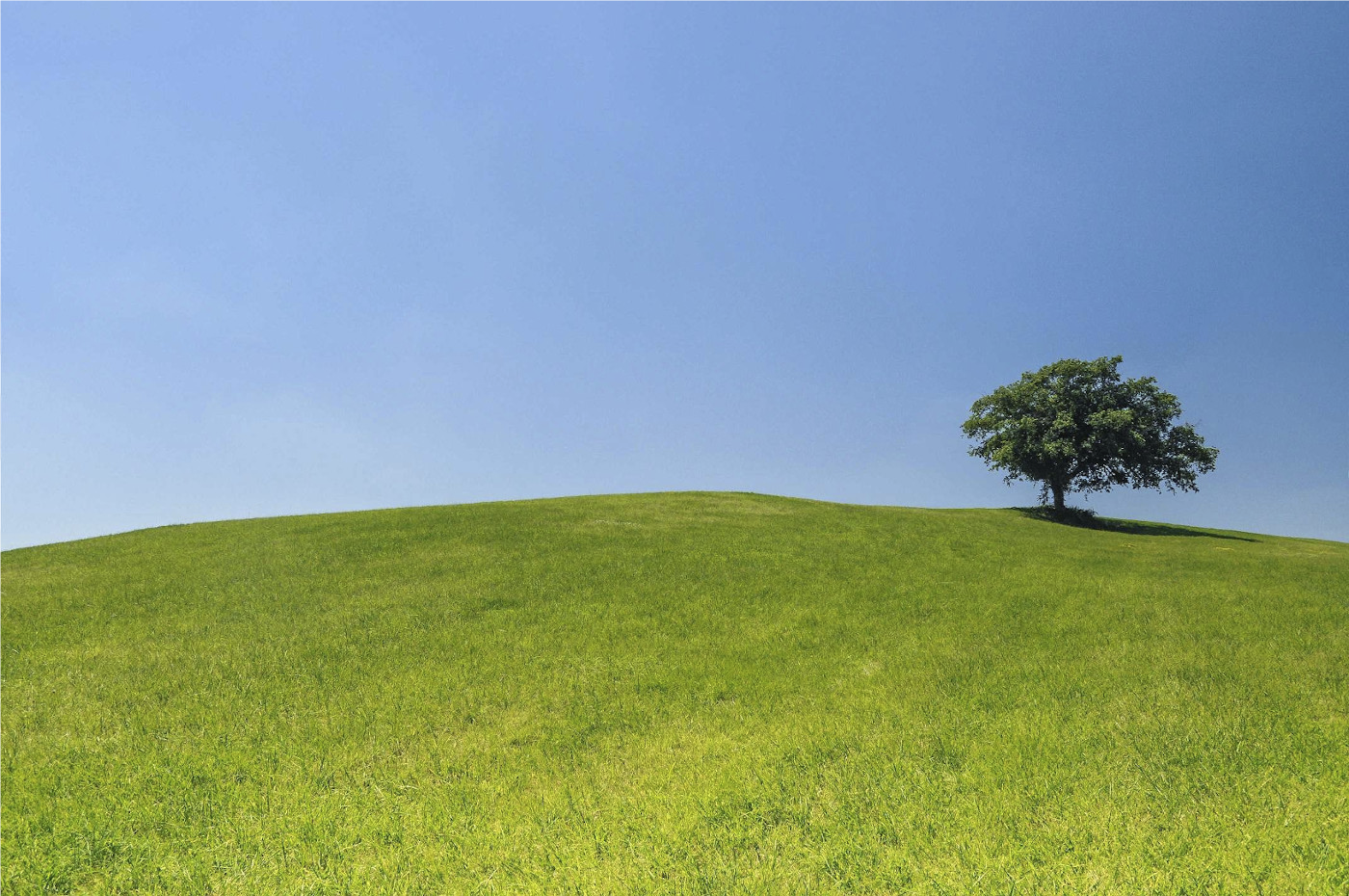}%
		\label{CART:Fig:task_3_in}%
	} \\
	\subfloat[Task-1 Output]{%
		\includegraphics[width=0.31\linewidth]{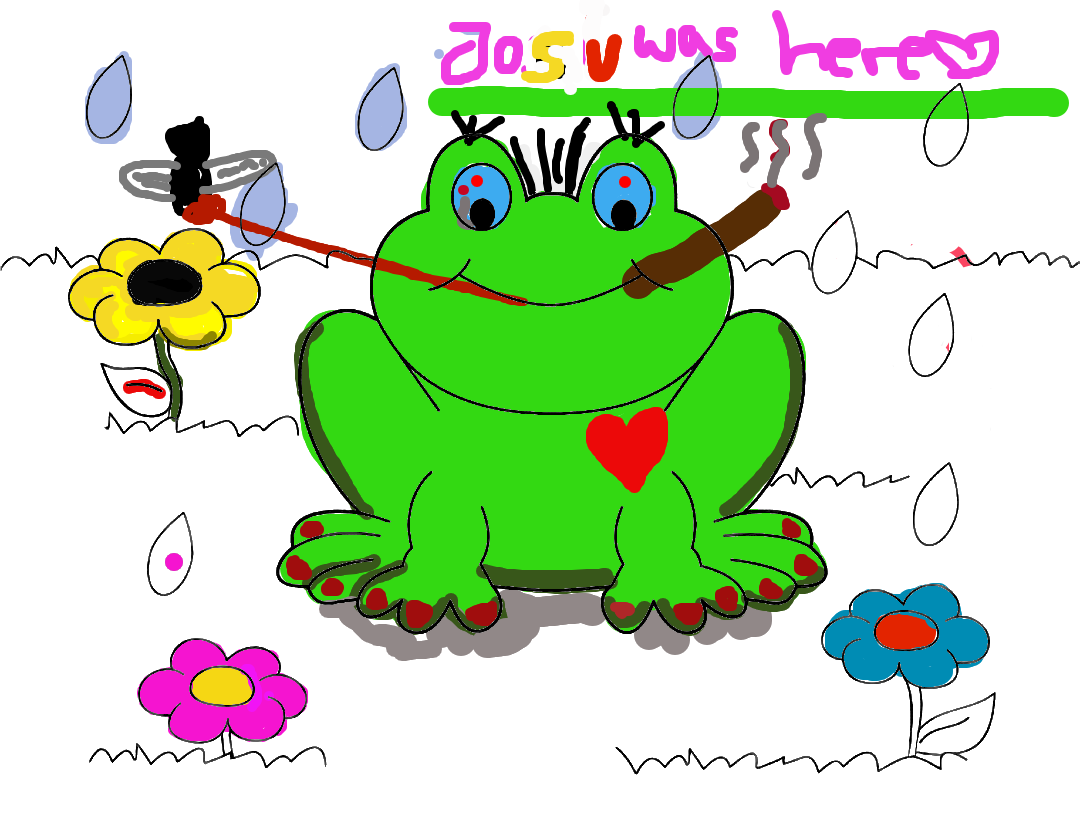}%
		\label{CART:Fig:task_1_out}%
	}\hfill
	\subfloat[Task-2 Output]{%
		\includegraphics[width=0.31\linewidth]{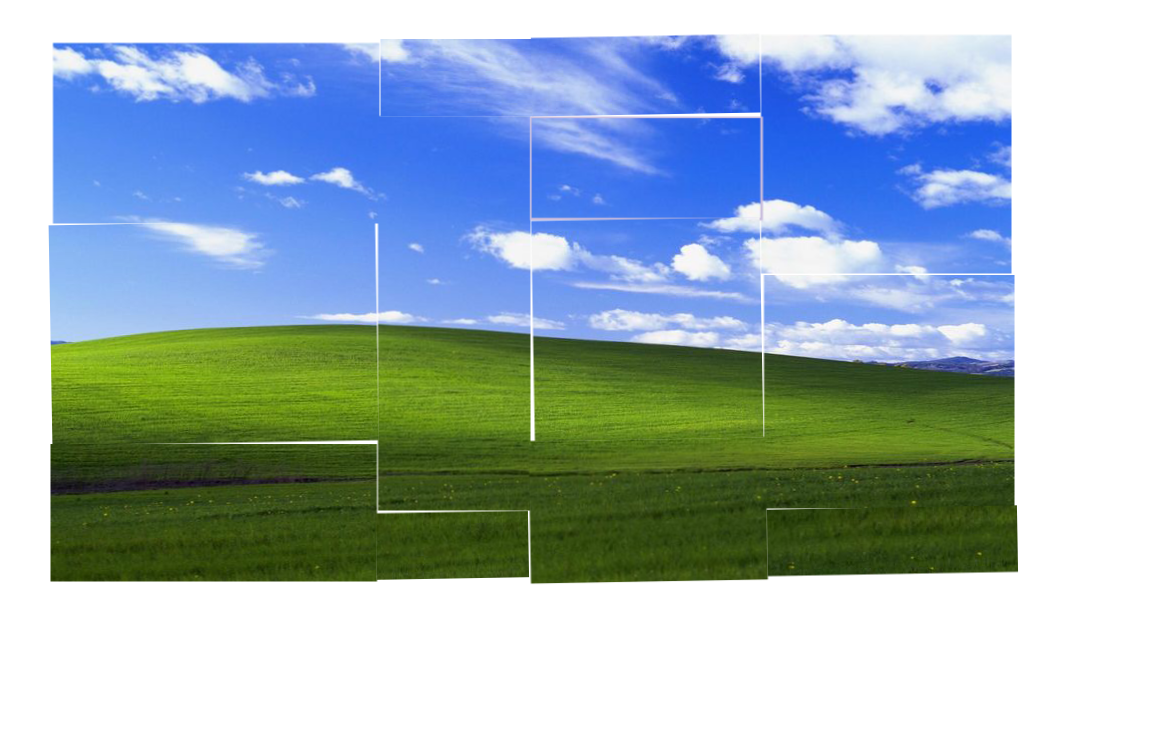}%
		\label{CART:Fig:task_2_out}%
	}\hfill
	\subfloat[Task-3 Output]{%
		\includegraphics[width=0.31\linewidth]{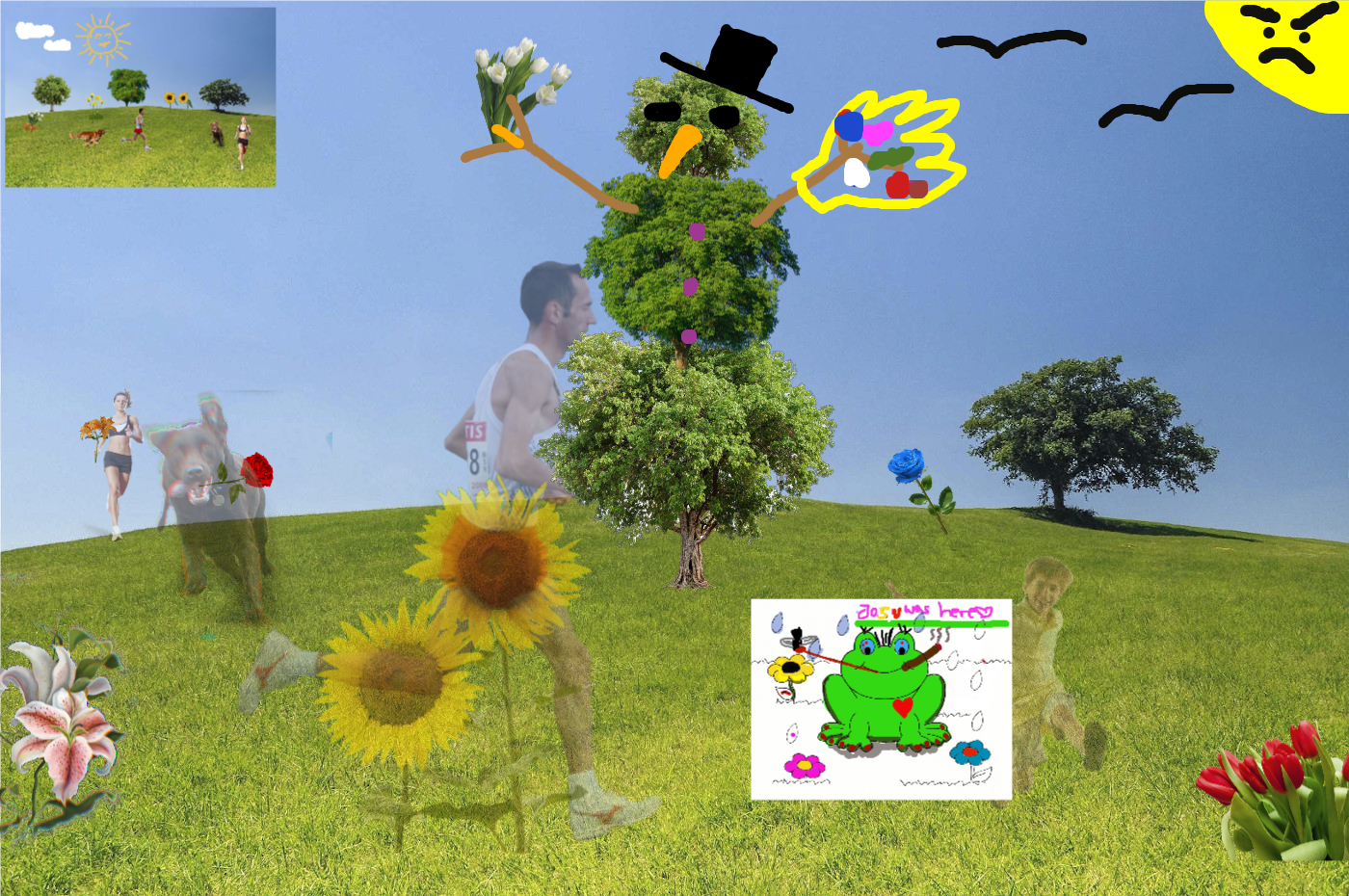}%
		\label{CART:Fig:task_3_out}%
	}
	\caption{Exemplary results obtained with our system during sessions of the post-deployment user study.}
	\label{CART:Fig:TaskResults}
	\belowfig	
\end{figure}

\subsection{User Tasks}

\noindent The three tasks performed by each participant group cover the full potential of our editing system.
The tasks are ordered by increasing difficulty and took \SIrange{15}{20}{\min} respectively for completion. 
\Cref{CART:Fig:TaskResults} shows selected results obtained during the study.

\paragraph{Sketching (Task-1)}
We provide a blank sketch as a background layer (\Cref{CART:Fig:task_1_in}) and the participants are asked to color the sketch using the brush tool on the empty top layer (\eg~\Cref{CART:Fig:task_1_out}). 
The users are encouraged to use multiple brush colors and also create their own doodle using an additional layer. 
The task objective is to test if users are able to work with layers, use the brush tool effectively, and detect potential synchronous conflicts.
We stopped this exercise once the users were familiar with the brush tool and working with layers; this task took \SIrange{10}{15}{\min}.

\paragraph{Puzzle (Task-2)}
We provide the users with a set of disarranged pieces of a test image~(\Cref{CART:Fig:task_2_in}). 
Each piece is represented in the form of a single layer. 
The task is to rearrange these layers using rotation and translation in order to solve the puzzle~(\Cref{CART:Fig:task_2_out}). 
The task objective is to test if users are able to use layer transformation tools effectively.
We also provide the puzzling image as a guide. 
On an average, it took between \SIrange{15}{20}{\min} to complete.

\paragraph{Collage Creation (Task-3)}
Given a set of images, \ie one background image and various foreground images with alpha matte (\Cref{CART:Fig:task_3_in}), the users should create and layout respective layers -- comprising as many foreground images as they like -- in order to create a collage collaboratively.
In addition thereto, they are encouraged to apply different image effects (using \acp{VCA} such as contrast enhancement, pixelation, chroma-zoom, or chromatic aberration) to each layer. 
The task objective is to test if users are able to reuse their learning from the previous tasks and also test familiarity with blending and layer modification via \acp{VCA}.
The time for this task was limited to \SI{15}{\min}.

\begin{figure}[tb]
	\centering
	\subfloat[Overall Satisfaction]{%
		\includegraphics[width=0.48\columnwidth]{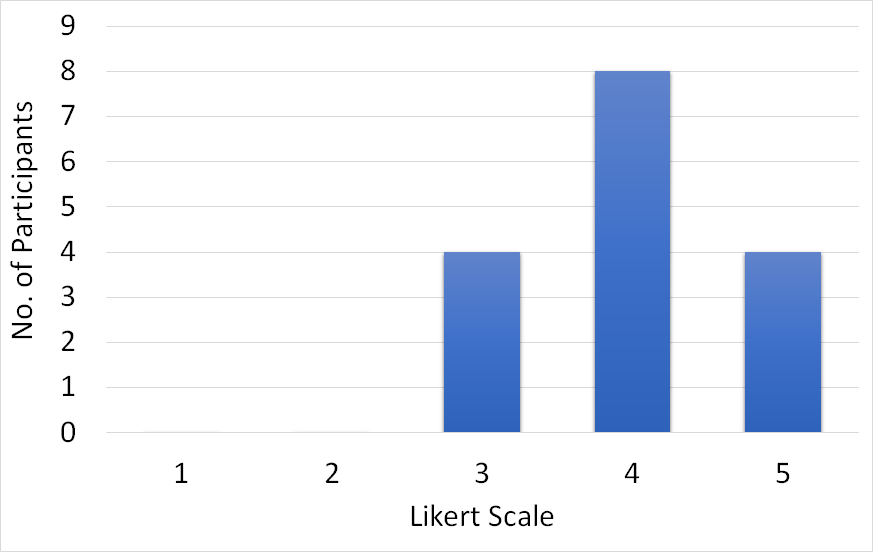}%
		\label{fig:post_deploy_a}%
	}\hfill
	\subfloat[Functional Satisfaction]{%
		\includegraphics[width=0.48\columnwidth]{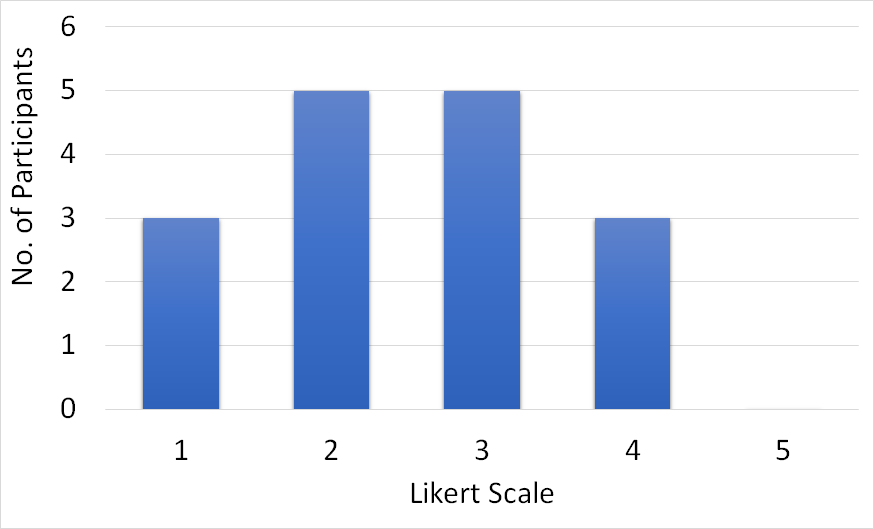}%
		\label{fig:post_deploy_b}%
	}
	\caption{The (a) overall and (b) functional satisfaction of the participants during the post-deployment user study on a Likert-Scale of \num{1} to \num{5}, with \num{5} being the best.}%
	\label{CART:Fig:PostDeployFeedback}
	\belowfig	
\end{figure}

\subsection{Data Collection and Analysis}
\noindent The online session of the above tasks is followed by a subjective interview (of approx. \SI{15}{\min}) with questions focusing on performance, collaboration, and potential applications. 
In addition thereto, the entire online session was video recorded to analyze groups' collaborative practices and also to record their feedback. 
After the interview, each participant is asked to file a post-study questionnaire based on QUIS and CSUQ without any time constraints.

All the participants were able to perform Task-1 quite easily and were satisfied with the system performance. 
It indicates that even in the current state our system can be used for a collaborative coloring-book application. 
For Task-2, the major difficulty was maintaining the control of a particular layer. 
Participants reported that the user-specific visual feedback regarding layer selection was too subtle.
Thus, it happened that two participants were trying to move the same layer and faced unexpected results.
However, in the subjective interview, they confirmed that such editing conflicts could have been avoided with the layer locking functionality. 
For Task-3, the major challenge was in terms of adding effects to layers, most of the users were not able to figure out this functionality on their own. 
Overall the user feedback can be summarized into the following two categories.

\paragraph{Collaboration}
As expected, the novel collaborative aspect of our system was appreciated by most of our participants~(\Cref{fig:post_deploy_a}). 
They showed a great interest in having this collaborative functionality integrated into the image editing tool of their choice.
Our participants from different background suggested a broader utility of our system in domains of engineering, architecture, teaching, entertainment, academia, etc., thus indicating a wide user base. 
However, further improvements for collaboration was suggested mainly in terms of ($i$) an integrated voice communication channel, ($ii$) hiding layers created by other team members, and ($iii$) functionality known from collaborative document editing, \eg tagging and commenting.   

\paragraph{Editing}
Our prototype does not offer all the editing functionality generally available in a common image-editing application.
Most of the participants who are familiar with such tools noticed the lack of such functionality~(\Cref{fig:post_deploy_b}), \eg an eraser tool, a flood fill tool, or selective layer manipulation (applying \acp{VCA} only on a selected region of a layer).
However, the integration of collaborative versions of these tools is supported by our architecture.

To answer the initial questions as part of the post-deployment user study: ($i$) the users understood the general structure of the \ac{GUI},
($ii$) the visualization metaphors, to avoid editing conflicts, were not intuitive in the beginning but were easy to use after guidance, and
($iii$) users were quite satisfied with our prototype, especially with respect to its collaborative nature.

\section{Conclusions}
\label{CART:Sec:SystemOverview}

\noindent In this paper, we designed and evaluated a web-based system for real-time collaborative editing of raster images. 
To the best of our knowledge, ours is the first system that provides such a wide variety of image-edits in a collaborative fashion. 
In order to better understand the needs for such a system, we conducted a preliminary user study. 
Our prototype leverages the power of WebGL for interactive browser-based rendering, while synchronization is maintained via WebSocket connections.
Our interface re-uses and extends \ac{GUI} concepts from common image-editing applications.
The post-deployment user study indicates a substantial demand for such a system.
As part of future work, we would like to address the existing limitations.

\section*{Acknowledgments}
\noindent This work was partially funded by the German Federal Ministry of Education and Research (BMBF) through grant 01IS1809 (\enquote{mdViPro}) and 01IS19006 (\enquote{KI-LAB-ITSE}) and the Research School on \enquote{Service-Oriented Systems Engineering} of the Hasso Plattner Institute.

\bibliographystyle{IEEEtran}
\bibliography{IEEEabrv,colier}

\begin{acronym}
\acro{GUI}{Graphical User Interface}
\acro{VCA}{Visual Computing Asset}
\acro{CGCE}{Common Graphics Collaborative Editing}
\acro{SPA}{Single-Page Application}
\acro{JSON}{JavaScript Object Notation}
\acro{ID}{Identifier}
\acro{CIE}{Collaborative Image Editing}
\end{acronym}

\end{document}